\documentclass[12pt]{article}
\usepackage{amssymb}
\usepackage{mathrsfs}
\begin{document}
\title{Nonabelian dualization of plane wave backgrounds}
\author{Ladislav Hlavat\'y and Miroslav Turek}
\maketitle
{Czech Technical University in Prague, Faculty of Nuclear Sciences
and Physical Engineering,  B\v rehov\'a 7, 115 19 Prague 1, Czech
Republic}
\begin{abstract}{We investigate plane--parallel wave metrics from the point of view of their
(Poisson--Lie) T--dualizability. For that purpose we reconstruct the
metrics as backgrounds of nonlinear sigma models on Lie groups. For
construction of dual backgrounds we use Drinfel'd doubles
 obtained from the isometry groups of the metrics.

We find dilaton fields that enable to satisfy the vanishing beta
equations for the duals of  the homogenous plane--parallel wave
metric. Torsion potentials or $B$--fields, invariant w.r.t. the
isometry group of Lobachevski plane waves are obtained by the
Drinfel'd double construction.

We show that a certain kind of plurality, different from the
(atomic) Poisson--Lie T--plurality, may exist in case that metrics
admit several isometry subgroups having the dimension of the
Riemannian manifold. An example of that are two different
backgrounds dual to the homogenous plane--parallel wave metric.}

\end{abstract}
Keywords: Sigma model, string duality, pp-wave background

\def\tuc{\bf}
\def\ba{\begin{eqnarray}}
\def\ea{\end{eqnarray}}
\def\be{\begin{equation}}
\def\ee{\end{equation}}
\def\lbl{\label}
\def \rf  {(\ref}

\def\eqn{equation}
\def\cond{condition}
\def\tfn{transformation}
\def\soln{solution}
\def\fn{function}
\def\sm{$\sigma$--model}
\def\pl{Poisson--Lie}
\def\pltfn{Poisson--Lie transformation}
\def\dd{Drinfel'd double}
\def\vbe{vanishing $\beta$ function equations}
\def\3dial{three--dimensional}
\def\-1{^{-1}}
\def\half{\frac{1}{2}}
\def\coor{coordinate}
\def\real{{\bf R}}
\def\compl{{\bf C}}

\def\e{{\rm e}}
\def\real{{\bf R}}
\def\cd{{\mathcal D}}
\def\cg{{\mathcal G}}
\def\tcg{\widetilde{{\mathcal G}}}
\def\tx{\widetilde{X}}

\def\th{\widetilde{h}}
\def\sm{$\sigma$--model}
\def\pltp{Poisson--Lie T--pluralit}
\def\pltd{Poisson--Lie T--dualit}
\def\dd{Drinfel'd double}
\section{Introduction}
Sigma models can serve as models of string theory in curved and
time-dependent backgrounds. Solution of sigma-models in such
backgrounds is often very complicated, not to say impossible. On the
other hand, there are many backgrounds whose properties were
thoroughly investigated and it is therefore interesting to find if
they can be transformed to some others. Important example of such
transformation is so called Poisson Lie T-duality.

In their seminal work \cite{klise} Klim\v c\'ik and \v Severa set
conditions for dualizability of backgrounds and gave formulas for
their transformation. Since then several examples of dualizable
sigma models were constructed, see e.g. \cite {vall:su2},
\cite{sfe:pltd}, \cite{hlas}. Unfortunately, most of the examples
are not
 physically interesting. The purpose
of this paper is to show that physical backgrounds that admit
sufficiently large group of isometries are naturally dualizable and
therefore equivalent in a sense to some others. In this paper we are
going to investigate four--dimensional plane--parallel wave metrics
\cite{papa,blau, siklos,podolsky} from this point of view.

The basic concept used for construction of dualizable sigma models
is \dd {} -- Lie group with additional structure. The Drinfel'd
double for a sigma model living in curved background can sometimes
be found from the knowledge of symmetry group of the metric. More
precisely, in the \dd{} there are two equally dimensional subgroups
whose Lie algebras are  isotropic subspaces of the Lie algebra of
the \dd. In case that the metric has sufficient number of
independent Killing vectors, the isometry group of the metric (or
its subgroup) can be taken as one of the subgroups of the \dd. The
other one then must be chosen abelian in order to satisfy the
conditions of dualizability. Short summary of the dualization
procedure described e.g. in \cite{kli:pltd} is given in the next
section.

\section{Elements of Poisson-Lie T-dual sigma-models}\label{secPLT}
Let $G$ be a Lie group and ${\cal{G}}$ its Lie algebra. Sigma model
on the group $G$ is given by the classical action
    \be\label{sma}
        S_{F}[\phi]=\int d^{2}\sigma\,\partial_{-}\phi^{\mu}F_{\mu\nu}(\phi)\partial_{+}\phi^{\nu},
    \ee
where $F$ is a second order tensor field on the Lie group $G$. The functions
$\phi^{\nu}$$:\mathbf{{R}}^{2}\rightarrow\mathbf{R}$, $\mu =1,2,\ldots,dim(G),$ are determined by the composition
$\phi^{\mu}=x^{\mu}\circ g$ where $g:{\mathbf{R}^{2}}\ni(\sigma_+,\sigma_-)\mapsto g(\sigma_+,\sigma_-)\in G$ and
$x^{\mu}:\mathbf{U}_{g}\rightarrow \mathbf{R}$ are components of a coordinate map of neighborhood $\mathbf{U}_{g}$ of
element $g(\sigma_+,\sigma_-)\in G$.

Equivalently the action can be expressed as
    \be
        S_{F}[g]=\int d^{2}x R_{-}(g)^{a}E_{ab}(g)R_{+}(g)^{b},
    \ee
where
$R_{\pm}$ are right-invariant fields
$R_{\pm}(g):=(\partial_{\pm}gg^{-1})^{a}T_{a}\in {\cal{G}}$. The
relationship between $E$ and $F$ is given by the formula
    \be\label{met}
        F_{\mu\nu}(x)=e_{\mu}^{a}(g(x))E_{ab}(g(x))e_{\nu}^{b}(g(x)),
    \ee
where $e_{\mu}^{a}(g(x)$ are the components of right invariant forms $e_{\mu}^{a}=((dg)g^{-1})_{\mu}^{a}$. The
equations of motion derived from the action $(\ref{sma})$ have the following form
    \be\label{eqnofmsigma}
        \partial_{-}\partial_{+}\phi^{\mu}+\Gamma_{\nu\lambda}^{\mu}\partial_{-}\phi^{\nu}\partial_{+} \phi^{\lambda}=0,
    \ee
where $\Gamma_{\nu\lambda}^{\mu}$ are components of the Levi-Civita
connection associated with the second order tensor field $F$ This
tensor field is a composition of the metric (a symmetric part) and
the torsion potential (an antisymmetric part). The condition of
dualizability of sigma-models on the level of  the Lagrangian is
given by the formula \cite{klise}
    \be\label{dc}
        {\cal{L}}_{v_{i}}F_{\mu\nu}=F_{\mu\kappa}v_{j}^{\kappa}\tilde{c}^{jk}_{i}v^{\lambda}_{k}F_{\lambda\nu},
    \ee
where $\tilde{c}_{i}^{jk}$ are structure coefficients of the dual
algebra ${\cal{\widetilde G}}$ and $v_{i}$ are left-invariant fields
on the Lie group $G$. The algebras ${\cal {G}}$ and
${\cal{\widetilde G}}$ then define the \dd{} that enables to
construct tensor $F$ satisfying (\ref{dc}).
\subsection{The Drinfel'd double and Poisson--Lie T--duality}
As mentioned in the Introduction the Drinfel'd double $D$ is defined
as a connected Lie group whose Lie algebra ${\cal{D}}$ can be
decomposed into pair of subalgebras ${\cal{G}},\tilde{\cal G}$
maximally isotropic with respect to a symmetric ad-invariant
nondegenarate bilinear form $<.,.>$ on ${\cal{D}}$.

Under the condition (\ref{dc}) the field equations
(\ref{eqnofmsigma}) for the \sm{} can be rewritten as \eqn{} for the
mapping $l(\sigma_+,\sigma_-)$ from the world-sheet $\mathbf{R}^2$
into the Drinfel'd double $D$
    \be \label{deqn}
        <(\partial_{\pm}l)l^{-1},\varepsilon^{\mp}>=0,
    \ee
where subspaces $\varepsilon^{+}=span(T^{i}+E^{ij}(e){\tilde
T}_{j})$, $\varepsilon^{-}=span(T^{i}-E^{ji}(e){\tilde T}_{j})$ are
orthogonal w.r.t. $<,>$ and span the whole Lie algebra $\cal D$.
$\{T^{i}\}$, $\{{\tilde T}_{j}\}$ are the bases of ${\cal{G}}$ and
$\tilde{\cal G}$.

Due to Drinfel'd, there exists unique decomposition (at least in the vicinity of the unit element of $D$) of an
arbitrary element $l$ of $D$ as a product of elements
from $G$ and $\tilde G$. The solutions of \eqn{} (\ref{deqn}) and
solution $\phi^\mu(\sigma_+,\sigma_-) = (x^\mu\circ
g)(\sigma_+,\sigma_-)$ of the \eqn(\ref{eqnofmsigma}) are related by
    \be
        l(\sigma_+,\sigma_-)=g(\sigma_+,\sigma_-){\tilde{h}}(\sigma_+,\sigma_-)\in D,
    \ee
where $g\in G$, ${\tilde {h}}\in {\tilde{G}}$ fulfil the equations
    \ba
        (\partial_{+}{\tilde{h}}\,{\tilde {h}}^{-1})_{a}&=& -(\partial_{+}g\,g^{-1})^{b}E_{cb}(g)d_{a}^{c}(g)\\
        (\partial_{-}{\tilde{h}}\,{\tilde {h}}^{-1})_{a}&=& (\partial_{-}g\,g^{-1})^{b}E_{bc}(g)d_{a}^{c}(g),
    \ea

The matrix $E(g)$ of the dualizable $\sigma$-model is of the form
    \be\label{metr}
        E(g)=[E_{0}^{-1}+\Pi(g)]^{-1},
    \ee
where $E_{0}$ is a constant matrix, $\Pi(g)$ is given by the formula
    \be
        \Pi(g)=b(g).a(g)^{-1}=-\Pi^{t}(g)
    \ee
and matrices $a(g),b(g),d(g)$ are given by the adjoint representation of the Lie subgroup $G$ on the Lie algebra of
the Drinfel'd double in the basis $\{T^{i},{\tilde T}_{j}\}$\footnote{The superscript $t$ means transposition of the
matrix}
    \be
        Ad(g)^{t}=\left(
                \begin{array}{cc}
                    a(g) & 0 \\
                    b(g) & d(g) \\
                \end{array}
      \right).
    \ee
Let us note that $E_0$ is the value of $E(g)$ in the unit $e$ of the
group $G$ because $b(e)=\Pi(e)=0$.

The dual model can be obtain by the exchange \be \label{duality}
G\leftrightarrow{\widetilde G},\ \ \ {\cg}\leftrightarrow
{\cal{\widetilde G}},\ \ \ \Pi(g)\leftrightarrow
{\widetilde{\Pi}(\tilde g)},\ \ \ E_{0}\leftrightarrow
E_{0}^{-1}.\ee Solutions of the equations of motion of dual models
are mutually associated by the relation
    \be
        l(\sigma_+,\sigma_-)= g(\sigma_+,\sigma_-)\tilde h(\sigma_+,\sigma_-)=\tilde g(\sigma_+,\sigma_-)h(\sigma_+,\sigma_-).
    \ee
\subsection{Poisson--Lie T--plurality} Generally, more than two decompositions\footnote{Two decompositions always exist,
$({\cal{G}}|{\tilde{\cal{G}}}),({\tilde{\cal{G}}}|{\cal{G}})$}
{\it{(Manin triples)}} of Lie algebra $\cd$ of the Drinfel'd double
can exist. This possibility leads to {\it{Poisson--Lie
T--plurality}}. Let $({\widehat{\cal{G}}}|{\overline{\cal{G}}})$ is
another decomposition of the Drinfel'd algebra
$\cd=({\cal{G}}|\tilde{\cal G})$ into a pair of maximal isotropic
subalgebras. Then the Poisson--Lie T--plural sigma model is given by
the following formulas \cite{unge:pltp}
    \ba\label{ptE}
    \widehat{E}(\hat{g})&=&(\widehat{E}_0^{-1}+\widehat{\Pi}(\hat{g}))^{-1}, \\
      \widehat{\Pi}(\hat{g})&=&\widehat{b}(\hat{g})\cdot \widehat{a}(\hat{g})^{-1} ,\nonumber\\
      \widehat{E}_0&=&(K+E_0\cdot R)^{-1}\cdot (Q+E_0\cdot S),
    \ea
where the matrices $K,Q,R,S$ determine the relationship between the
bases of the appropriate decompositions
${\cal{G}},{\tilde{\cal{G}}}$ and
${\widehat{\cal{G}}},{\overline{\cal{G}}}$
    \be
         \left(
                \begin{array}{cc}
                    T\\
                    \widetilde{T}
                \end{array}
            \right)=
            \left(
                \begin{array}{cc}
                    K & Q \\
                    R & S \\
                \end{array}
            \right)
            \left(
              \begin{array}{c}
                \widehat{T} \\
                \overline{T} \\
              \end{array}
            \right).
    \ee
The relationship between the classical solutions of the two
Poisson--Lie T-plural sigma-models is given by a possibility of two
decompositions of the element $l\in D$ as
    \be
          l(\sigma_+,\sigma_-)= g(\sigma_+,\sigma_-)  \tilde h(\sigma_+,\sigma_-)=\hat g(\sigma_+,\sigma_-) \bar h(\sigma_+,\sigma_-).
    \ee
The Poisson--Lie T--duality is then a special case of \pl{}
T--plurality for $K=S=0,\ Q=R=1$.

\section{Homogenous plane wave metrics}
Homogenous plane wave is generally defined by the metric of the
following form \cite{papa},\cite{blau}
    \be \label{homogeneousmetric}
        ds^{2}=2dudv-A_{ij}(u)x^{i}x^{j}du^{2}+dx^{2},
    \ee
where $dx^{2}$ is the standard metrics on Euclidean space ${\bf{E}}^{d}$ and $x\in{\bf{E}}^{d}.$ The form of this
metric seems to be simple, but explicit construction of sigma models can be very complicated. Therefore, we have
focused on the special case of isotropic homogenous plane wave metric $A_{ij}(u)=\lambda(u)\delta_{ij}$
    \be\label{hrvm}
        ds^{2}=2dudv-\lambda(u)x^2 du^{2}+dx^{2}.
    \ee
Metric $(\ref{hrvm})$ has a number of symmetries important for the
construction of the dualizable sigma models. It admits the following
Killing vectors
    \ba
        T&=&\partial_{v}\\\nonumber
        X_{i}&=&a\partial_{i}-\partial_{u}ax_{i}\partial_{v}\\\nonumber
        R_{ij}&=&x_{i}\partial_{j}-x_{j}\partial_{i},
    \ea
where $a(u)$ satisfies
    \be\label{efa}
        \partial_{u}^{2} a+\lambda a=0.
    \ee
The Killing vectors $R_{ij}$ are generators of orthogonal rotations
in ${\bf{E}}^{d}$.  For  special choice of
    \be \label{lambda}
        \lambda(u)=\frac{k}{u^{2}}, \quad k=const.
    \ee
there are further isometries related to the scaling of the
light-cone coordinates
    \be
        u\rightarrow \kappa\, u,\quad v\rightarrow \kappa^{-1}v.
    \ee
The specific form of $\lambda$ enables us to calculate the function
$a(u)$ explicitly.  The Killing vectors of the metric (\ref{hrvm})
for $ \lambda(u)=\frac{k}{u^{2}}$ are
    \ba
        T&=&\partial_{v}\\\nonumber
        X_{i}&=&u^{\nu}\partial_{i}-\nu
        u^{\nu-1}x_{i}\partial_{v}\\\nonumber
        {\tilde{X}}_{i}&=&u^{1-\nu}\partial_{i}-(1-\nu)u^{-\nu}x_{i}\partial_{v}\\\nonumber
        D&=&u\partial_{u}-v\partial_{v}\\\nonumber
        R_{ij}&=&x_{i}\partial_{j}-x_{j}\partial_{i},
    \ea
where $D$ is the generator associated with the scaling symmetry and
$k=\nu-\nu^{2}$.

In the following we shall investigate the case $d=2$. It means that the metric tensor in coordinates $(u,v,x,y)$ reads
    \be\label{FF}
        G_{ij}(u,v,x,y)=\left(
          \begin{array}{cccc}
            \frac{-k(x^{2}+y^{2})}{u^{2}} & 1 & 0 & 0 \\
            1 & 0 & 0 & 0 \\
            0 & 0 & 1 & 0 \\
            0 & 0 & 0 & 1 \\
          \end{array}
        \right).
    \ee
This metric is not flat but its Gaussian curvature vanishes. Note that it has singularity in $ u=0$. It does not
satisfy the Einstein \eqn s but the conformal invariance conditions equations for vanishing of the $\beta$-function
\begin{eqnarray}
\label{bt1} 0 & = & R_{ij}-\bigtriangledown_i\bigtriangledown_j\Phi-
\frac{1}{4}H_{imn}H_j^{mn}\,,
\\
 \label{bt2} 0 & = & \bigtriangledown^k\Phi H_{kij}+\bigtriangledown^k H_{kij}\,,
\\
\label{bt3} 0 & = & R-2\bigtriangledown_k\bigtriangledown^k\Phi-
\bigtriangledown_k\Phi\bigtriangledown^k\Phi-
\frac{1}{12}H_{kmn}H^{kmn}
\end{eqnarray}
where the covariant derivatives $\bigtriangledown_k$, Ricci tensor
$R_{ij}$ and Gauss curvature $R$ are calculated from the metric
$G_{ij}$ that is also used for lowering and raising indices. Torsion
$H$ in this case vanishes and dilaton field is \cite{papa}
\begin{equation}\label{dilatonhomogeneous} \Phi= \Phi_0 - c\,u+
2\nu(\nu-1)\ln\, u.
\end{equation}
The metric (\ref{FF}) admits the following Killing vectors\footnote{ If $\nu=1/2$, i.e. $k=1/4$ then $K_{2}=K_{4},\
K_{3}=K_{5}$ }
   \ba
\label{killingshomogeneous}
        K_{1}&=&\partial_{v}\\\nonumber
        K_{2}&=&u^{\nu}\partial_{x}-\nu
        u^{\nu-1}x\partial_{v}\\\nonumber
        K_{3}&=&u^{\nu}\partial_{y}-\nu
        u^{\nu-1}y\partial_{v}\\\nonumber
        K_{4}&=&u^{1-\nu}\partial_{x}-(1-\nu)u^{-\nu}x\partial_{v}\\\nonumber
        K_{5}&=&u^{1-\nu}\partial_{y}-(1-\nu)u^{-\nu}y\partial_{v}\\\nonumber
        K_{6}&=&u\partial_{u}-v\partial_{v}\\\nonumber
        K_{7}&=&x\partial_{y}-y\partial_{x}
        \ea
One can easily check that the Lie algebra spanned by these vectors is the semidirect sum $  {\cal S}\ltimes {\cal N}$
where ${\cal S}=Span[K_6,K_7]$ and ideal ${\cal N}=Span[K_1,K_2,K_3,K_4,K_5]$. The algebra ${\cal S}$ is abelian and
its generators can be interpreted as dilation in $u,v$ and rotation in $x,y$. Generators of the algebra ${\cal N}$
commute as two--dimensional Heisenberg algebra with the center $K_1$.
\subsection{Construction of dual metrics}\label{homogeneous} As explained in Section \ref{secPLT}, dualizable metric
can be constructed by virtue of \dd. For this goal the Lie algebra
$\cd$ of the Drinfel'd double can be composed from the
four-dimensional Lie subalgebra $\cal G$ isomorphic to the
four--dimensional subalgebra of Killing vectors  and
four-dimensional Abelian algebra\footnote{It is easy to see, that
the equation $(\ref{dc})$ is then fulfilled.} $\tilde{\cal G}$.
Moreover, the four--dimensional subgroup of isometries must act
freely and transitively \cite{klise} on the Riemannian manifold $M$
where the metric (\ref{hrvm},\ref{lambda}) is defined so that
$M\approx G$.

Using the method described in \cite{winter} for semisimple algebras
we find that up to the \tfn{} $\nu\mapsto 1-\nu$, i.e. $k\mapsto k$
there are six classes of four--dimensional subalgebras of the
isometry algebra of the homogeneous plane wave metric isomorphic to
\begin{itemize}
    \item $ Span[K_1,K_2+\rho K_5,K_3+\sigma K_5,K_4+\tau K_5]$
    \item $ Span[K_1,K_2,K_3,K_7]$
    \item $ Span[K_1,K_2,K_3,K_6+\rho\, K_7]$
    \item $ Span[K_1,K_2,K_4,K_6]$
    \item $ Span[K_1,K_2,K_5,K_6]$
    \item $ Span[K_2,K_3,K_6+\rho K_1,K_7]$
\end{itemize} where $\rho,\sigma,\tau $ are arbitrary parameters.

Infinitesimal form of transitivity condition can be formulated as
requirement that four independent Killing vectors  can be taken as
basis vectors of four--dimensional vector distribution in  $M$. In
other words, these Killing vectors must form a basis of tangent
space in every point of $M$. It means that in every point of $M$
there is an invertible matrix $A(u,v,x,y)$ that solves the \eqn
\begin{equation}\label{transit}
   \partial_\alpha=A_\alpha^\beta(u,v,x,y)\,X_\beta,
\end{equation}
where $\alpha,\beta=1,2,3,4$,
$\partial_\alpha=\partial_u,\partial_v,\partial_x,\partial_y$ and
$X_\beta$ form a basis of the subalgebra.

Infinitesimal form of requirement that the action of the isometry
subgroup is free says that if in any point of $M$ there is a vector
of the corresponding Lie subalgebra such that its action on the
point vanishes then it must be null vector.

By inspection we can find that the only  four--dimensional
 subalgebras that generate transitive actions on $M$
are isomorphic to $ Span[K_1,K_2,K_3,K_6+\rho\, K_7]$ or $
Span[K_1,K_2,K_5,K_6]$. Their non-vanishing commutation relations
are
    \ba\label{cr1}
       [K_{6}+\rho K_7,K_{1}]&=&K_{1},\\  \nonumber
        [K_{6}+\rho K_7,K_{2}]&=&\nu\, K_{2}-\rho\ K_3,\\ \nonumber
        [K_{6}+\rho K_7,K_{3}]&=&\nu\, K_{3}+\rho\ K_2,
    \ea
and
    \ba\label{cr2}
        [K_{6},K_{1}]&=&K_{1},\\ \nonumber
        [K_{6},K_{2}]&=&\nu\, K_{2},\\ \nonumber
        [K_{6},K_{5}]&=&(1-\nu)\, K_{5},
    \ea
respectively where $\nu$ and $\rho$ are real parameters. One can
also check that the action of both corresponding groups of
isometries is free. In the following we shall find metric dual to
(\ref{FF}) that follows from its \dd {} description {} where $\cal
G$ is isomorphic either to algebra spanned by
$(K_{1},K_{2},K_{3},K_{6}+\rho K_7)$ or by
$(K_{1},K_{2},K_{5},K_{6})$.

 Let us start with construction of the
\dd {} following from the algebra isomorphic to (\ref{cr1}) and dual
Abelian algebra. Assume that the Lie algebra ${\cal{G}}$ is spanned
by elements $X_{1},X_{2},X_{3},X_{4}$ with commutation relations
    \ba
        [X_{4},X_{1}]&=&X_{1},\\\nonumber
        [X_{4},X_{2}]&=&\nu X_{2}-\rho\ X_3,\\\nonumber
        [X_{4},X_{3}]&=&\nu X_{3}+\rho\ X_2,   \ea
where $\nu$ and $\rho$ are arbitrary real parameters. The basis of
left-invariant vector fields of the group generated by ${\cal{G}}$
is
\[e^{{x_4}} \frac{\partial}{\partial x_1},\]
 \be\label{rivf1}e^{\nu  {x_4}} \cos (\rho {x_4})\frac{\partial}{\partial x_2}
-e^{\nu  {x_4}} \sin (\rho {x_4}) \frac{\partial}{\partial
 x_3},\ee
\[ e^{\nu  {x_4}} \sin (\rho {x_4})\frac{\partial}{\partial x_2}+
e^{\nu  {x_4}} \cos (\rho {x_4})\frac{\partial}{\partial
 x_3},\]
\[ \frac{\partial}{\partial x_4},
 \] where $x_{1},x_{2},x_{3},x_{4}$ are group coordinates
used in parametrization
    \be
        g=e^{x_{1}X_{1}}e^{x_{2}X_{2}}e^{x_{3}X_{3}}e^{x_{4}X_{4}}.
    \ee
To be able to obtain the metric (\ref{FF}) by the \dd {}
construction first we have to transform it into the group
coordinates. Transformation between group coordinates
$x_{1},x_{2},x_{3},x_{4}$ and geometrical coordinates $u,v,x,y$ is
     \ba \label{tfncoorshomogeneous}
        u&=&e^{x_{4}}\\\nonumber
        v&=&[-\frac{1}{2}\nu(x_{2}^{2}+x_{3}^{2})+x_{1}]e^{-x_{4}}\\\nonumber
        x&=&x_{2}\cos(\rho\, x_4)-x_3\sin(\rho\, x_4)\\\nonumber
        y&=&x_{3}\cos(\rho\, x_4)+x_2\sin(\rho\, x_4).
    \ea
It converts the Killing vectors $K_{1},K_{2},K_{3},K_{6}+\rho K_7$
into the left-invariant vector fields $(\ref{rivf1})$ and the metric
$(\ref{FF})$ into the form
    \be\label{F}
        F_{ij}(x_{1},x_{2},x_{3},x_{4})=\left(
                     \begin{array}{cccc}
                       0 & 0 & 0 & 1\\
                      0 & 1 & 0 & -\nu x_{2}-\rho\, x_3\\
                       0 & 0 & 1 & -\nu x_{3}+\rho\, x_2 \\
                       1 & -\nu x_{2}-\rho\, x_3 & -\nu x_{3}+\rho\, x_2 & -2x_{1}+(\nu^{2}+\rho^2)(x_{2}^{2}+x_{3}^{2}) \\
                     \end{array}
                   \right).
    \ee
that is obtainable by (\ref{met}) and (\ref{metr}). To get the
matrix $E_0$ necessary for construction of the dual model we note
that it is given by the value of $E(g)$ in the unit of the group,
i.e. by value of $ F_{ij}$ for $x_{1}=x_{2}=x_{3}=x_{4}=0$.
    \be \label{E0homogeneous}
        E_{0}=\left(
            \begin{array}{cccc}
                0 & 0 & 0 & 1\\
                0 & 1 & 0 & 0\\
                0 & 0 & 1 & 0\\
                1 & 0 & 0 & 0\\
            \end{array}
        \right).
    \ee

The dual tensor on the Abelian group $\tilde G$ constructed by the
procedure explained in the Section \ref{secPLT}, namely by using
(\ref{met}),(\ref{metr}) and (\ref{duality}) is
\begin{equation}\label{dualhomogenoeous}
\widetilde  F_{ij}(\tilde{x})=   \left(
\begin{array}{cccc}
 \frac{(\nu ^2+\rho^2) \left(\tilde{x_2}^2+\tilde{x_3}^2\right)}{\tilde{x_1}^2-1} & \frac{\nu\,  \tilde{x_2}-\rho\,\tilde{x_3}}{1-\tilde{x_1}} &
 \frac{\nu\,
   \tilde{x_3}+\rho\,\tilde{x_2}}{1-\tilde{x_1}} & \frac{1}{1-\tilde{x_1}} \\
 \frac{-\nu\,  \tilde{x_2}+\rho\,\tilde{x_3}}{\tilde{x_1}+1} & 1 & 0 & 0 \\
 \frac{-\nu\,  \tilde{x_3}-\rho\,\tilde{x_2}}{\tilde{x_1}+1} & 0 & 1 & 0 \\
 \frac{1}{\tilde{x_1}+1} & 0 & 0 & 0
\end{array}
\right).
\end{equation}One can see that the dual tensor 
has also antisymmetric part ($\widetilde B$--field or torsion
potential) \be \widetilde  B_{ij}=\frac{1}{2}(\widetilde
F_{ij}-\widetilde F_{ji})\,. \label{torpot} \ee
 and its torsion
$\widetilde  H=d\widetilde  B$  is
\begin{equation}\label{dualtorsionhom}
  \widetilde   H= \frac{2\rho}{\tilde{x_1}^2-1}\,d\tilde{x_1}\wedge d\tilde{x_2}\wedge d\tilde{x_3}.
\end{equation}

 The Gauss curvature of its symmetric part vanishes but the Ricci
tensor is nontrivial.  Dual metric that is symmetric part of
(\ref{dualhomogenoeous}) does not solve the Einstein \eqn s either
but again we can satisfy conformal invariance conditions
(\ref{bt1})--(\ref{bt3})  by the dilaton field
\begin{equation}\label{dualdilatonhomogeneous2} \widetilde  \Phi= \widetilde  \Phi_0 +
C \ln\left(\frac{\tilde{x_1}-1}{\tilde{x_1}+1}\right)
-\nu(\nu+1)\ln(\tilde{x_1}^2-1).
\end{equation}

If we use the subalgebra of isometries  spanned by
$(K_{1},K_{2},K_{5},K_{6})$ instead of that spanned by
$(K_{1},K_{2},K_{3},K_{6}+\rho K_7)$ then the transformation between
group coordinates $x_{1},x_{2},x_{3},x_{4}$ and geometrical
coordinates $u,v,x,y$ is \ba \label{tfncoorshomogeneous2}
        u&=&e^{x_{4}}\\\nonumber
        v&=&\frac{1}{2}\,[2x_{1}-\nu(x_{2}^{2}-x_{3}^{2})-x_{3}^{2}]e^{-x_{4}}\\\nonumber
        x&=&x_{2}\\\nonumber
        y&=&x_{3},
    \ea
the matrix $E_0$ gets again the form (\ref{E0homogeneous}) and
we get another tensor dual to (\ref{FF})
\begin{equation}\label{dualhomogenoeous2}
\widetilde  F_{ij}(\tilde{x})=   \left(
\begin{array}{cccc}
 \frac{\nu ^2 \tilde{x_2}^2+\left(1-\nu\right)^2\tilde{x_3}^2}{\tilde{x_1}^2-1} & \frac{\nu\,  \tilde{x_2}}{1-\tilde{x_1}} &
 \frac{\left(\nu-1\right)\,
   \tilde{x_3}}{1-\tilde{x_1}} & \frac{1}{1-\tilde{x_1}} \\
 -\frac{\nu\,  \tilde{x_2}}{\tilde{x_1}+1} & 1 & 0 & 0 \\
 \frac{\left(\nu-1\right)\,  \tilde{x_3}}{\tilde{x_1}+1} & 0 & 1 & 0 \\
 \frac{1}{\tilde{x_1}+1} & 0 & 0 & 0
\end{array}
\right).
\end{equation}
Even though it is not symmetric its torsion is zero. It satisfies
the conformal invariance conditions (\ref{bt1})--(\ref{bt3}) with
the dilaton field
\begin{equation}\label{dualdilatonhomogeneous3} \widetilde \Phi= \widetilde \Phi_0 +
C \ln\left(\frac{\tilde{x_1}-1}{\tilde{x_1}+1}\right)
+(\nu-1-\nu^2)\ln(\tilde{x_1}^2-1).
\end{equation}

\section{Lobachevsky plane waves}Another type of metrics that have
rather large group of isometries are so called Lobachevsky plane
waves \cite{siklos,podolsky}. They are of general form
\begin{equation}\label{siklosmetric}
  G_{ij}(u,v,x,y)=  \left(
\begin{array}{cccc}
 -\frac{H(u,x,y)}{b^2 x^2} & -\frac{1}{b^2 x^2} & 0 & 0 \\
 -\frac{1}{b^2 x^2} & 0 & 0 & 0 \\
 0 & 0 & -\frac{1}{b^2 x^2} & 0 \\
 0 & 0 & 0 & -\frac{1}{b^2 x^2}
\end{array}
\right)
\end{equation}
They satisfy Einstein equation with cosmological constant $3b^2$ iff
\begin{equation}\label{einsteincond}
    \frac{\partial^2}{\partial y^2} H(u,x,y)-\frac{2}{x}
   \frac{\partial}{\partial x}{H(u,x,y)}+
 \frac {\partial^2}{\partial x^2} H(u,x,y) =0
\end{equation}
The Gauss curvature of this metric is $-12\,b^2$. For special forms
of function $H$ the metric (\ref{siklosmetric}) admits various sets
of Killing vectors. All of them are subalgebras of a vector space
spanned by
\begin{eqnarray}\label{killings}\nonumber  K_I &=& \frac{\partial}{\partial v} \\
\nonumber  K_{II} &=& u\,\frac{\partial}{\partial u}-v\,\frac{\partial}{\partial v} \\
\nonumber  K_{III} &=& \frac{\partial}{\partial u} \\
  K_{IV} &=& \frac{\partial}{\partial y} \\
\nonumber  K_V &=& y\,\frac{\partial}{\partial
v}-u\,\frac{\partial}{\partial y}
 \\
\nonumber  K_{VI} &=& (2-\alpha)\,u\,\frac{\partial}{\partial
u}+(2+\alpha)\,v\,\frac{\partial}{\partial
v}+2x\,\frac{\partial}{\partial x}+2y\,\frac{\partial}{\partial y}\\
\nonumber  K_{VIII} &=& u^2\,\frac{\partial}{\partial
u}-\frac{1}{2}(x^2+y^2),\frac{\partial}{\partial
v}+ux\,\frac{\partial}{\partial x}+uy\,\frac{\partial}{\partial y}
\end{eqnarray}
A bit surprisingly, all these seven independent vector fields found
in \cite{siklos} form a Lie algebra even though they are not Killing
vectors of the same metrics (it depends on the form of $H(u,x,y)$).
We are interested in metrics that admit at least four independent
Killing vectors because they can be interpreted as dualizable
backgrounds for sigma models in four dimensions.

As mentioned in the Section \ref{homogeneous}, for construction of dualizable metrics we need a four--dimensional
subalgebra of Killing vectors that generates group of isometries that acts freely and transitively on the
four--dimensional Riemannian manifolds. Here we shall investigate metrics of the form (\ref{siklosmetric}) where that
{$H= x^\alpha$}, i.e
\begin{equation}\label{metricXtoalpha}
 G_{ij}(u,v,x,y)=    \left(
\begin{array}{cccc}
 -\frac{x^{\alpha -2}}{b^2} & -\frac{1}{b^2 x^2} & 0 & 0 \\
 -\frac{1}{b^2 x^2} & 0 & 0 & 0 \\
 0 & 0 & -\frac{1}{b^2 x^2} & 0 \\
 0 & 0 & 0 & -\frac{1}{b^2 x^2}
\end{array}
\right).
\end{equation}
It solves the Einstein equation with the cosmological constant $3b^2$ for $\alpha=3$ \cite{kaigor}.

\subsection{Construction of the dual metric}
The metric (\ref{metricXtoalpha}) has five--dimensional Lie group of isometries generated by the Killing vectors
$K_I,K_{III},K_{IV},K_V,K_{VI}$. Their nonzero commutators read
\begin{eqnarray}\label{comrelnsHxu}
\nonumber [K_I,K_{VI}] &=& (2+\alpha)\,K_I, \\
\nonumber  [K_{III},K_{V}] &=& -K_{IV}, \\
 \left[K_{III},K_{VI}\right] &=& (2-\alpha)\,K_{III}, \\
\nonumber   [K_{IV},K_{V}] &=& K_{I}, \\ \nonumber [K_{IV},K_{VI}] &=& 2\,K_{IV}, \\ \nonumber [K_{V},K_{VI}] &=&
\alpha\,K_{V}.
\end{eqnarray}

Four--dimensional subalgebras of the Lie algebra (\ref{comrelnsHxu}) for generic $\alpha$ are isomorphic to one of the
following algebras:
\begin{itemize}
    \item $ Span[K_I,K_{III},K_{IV},K_{VI}+\beta\,K_{V}]$
     \item $ Span[K_I,K_{III},K_{IV},K_V]$
    \item $ Span[K_I,K_{IV},K_{V},\delta\,K_{III}+\gamma\, K_{VI}]$
    \end{itemize}
It is easy to check that the only subalgebra of these that satisfy
the condition of transitivity (\ref{transit}) in every point of $M$
is the first one. Its action is free on $M$ as well so that we can
use it  for dualization of the metric (\ref{metricXtoalpha}).

In the following we shall consider the case $\beta=0$ because
$\beta\neq 0$ do not bring anything qualitatively different. It
means that for dualization we shall use the algebra $\cal G$ spanned
by $K_I,\,K_{III},\,K_{IV},\,K_{VI}$ with nonzero commutation
relations
\begin{eqnarray}\label{comrelnsXtoalpha}
 [K_I,K_{VI}] &=& (2+\alpha)\,K_I, \\
\nonumber  [K_{III},K_{VI}] &=& (2-\alpha)\,K_{III}, \\
\nonumber   [K_{IV},K_{VI}] &=& 2\,K_{IV}.
\end{eqnarray}

The corresponding Drinfel'd double is generated by the algebra $\cal
G$ defined by the commutation relations (\ref{comrelnsXtoalpha}) and
four--dimensional Abelian algebra. The basis of left--invariant
vector fields of the group generated by $\cal G$ is
\begin{equation}\label{rinvversXtoalpha}
 e^{- (2+\alpha )\,x_4}\frac{\partial}{\partial x_1} ,\
 e^{- (2-\alpha )\,x_4}\frac{\partial}{\partial x_2},\
e^{-{2x_4} } \frac{\partial}{\partial x_3} ,\
\frac{\partial}{\partial x_4}
\end{equation}
where $x_1,\,x_2,\,x_3,\,x_4$ are group coordinates used in parametrization
$$g(x_1,\,x_2,\,x_3,\,x_4)= e^{x_1X_1}e^{x_2X_2}e^{x_3X_3}e^{x_4X_4}$$
and $X_1,\,X_2,\,X_3,\,X_4$ are generators of $\cal G$ satisfying
\begin{eqnarray}\label{alraXtoalpha}
 [X_1,X_4] &=& (2+\alpha)\,X_1, \\
\nonumber  [X_2,X_4] &=& (2-\alpha)\,X_2, \\
\nonumber  [X_3,X_4] &=& 2\,X_3.
\end{eqnarray}

Transformation between group coordinates and coordinates $u,v,x,y$
of the Lobachevsky manifold is
\begin{eqnarray}\label{tfnCoorsXtoalpha}
  x_1 &=& v\,x^{-1-\frac{\alpha}{2}}, \nonumber\\
  x_2 &=& u\,x^{-1+\frac{\alpha}{2}}, \nonumber\\
  x_3 &=& \frac{y}{x}, \\
  x_4 &=& \frac{1}{2}\,\ln(x). \nonumber
\end{eqnarray}
This transformation converts the Killing vectors
$K_I,\,K_{III},\,K_{IV},\,K_{VI}$ into the left--invariant vector
fields (\ref{rinvversXtoalpha}) and the metric
(\ref{metricXtoalpha}) into
$$F_{ij}(x_1,\,x_2,\,x_3,\,x_4) =\left(
\begin{array}{cccc}
 0 & -\frac{1}{b^2} & 0 & \frac{{x_2} (\alpha -2)}{b^2} \\
 -\frac{1}{b^2} & -\frac{1}{b^2} & 0 & \frac{{x_2} (\alpha -2)-{x_1} (\alpha +2)}{b^2} \\
 0 & 0 & -\frac{1}{b^2} & -\frac{2 {x_3}}{b^2} \\
 \frac{{x_2} (\alpha -2)}{b^2} & \frac{{x_2} (\alpha -2)-{x_1} (\alpha +2)}{b^2} &
   -\frac{2 {x_3}}{b^2} & \frac{-{x_2}^2 (\alpha -2)^2-4 \left({x_3}^2+1\right)+2
   {x_1} {x_2} \left(\alpha ^2-4\right)}{b^2}
\end{array}
\right).$$ The value of this metric for $x_1=x_2=x_3=x_4=0$, i.e. in the unit of the group, gives the matrix
\begin{equation}\label{E0Xtoalpha}
    E_0=\left(
\begin{array}{cccc}
 0 & -\frac{1}{b^2} & 0 & 0 \\
 -\frac{1}{b^2} & -\frac{1}{b^2} & 0 & 0 \\
 0 & 0 & -\frac{1}{b^2} & 0 \\
 0 & 0 & 0 & -\frac{4}{b^2}
\end{array}
\right)
\end{equation}
Having this matrix we can construct the dual tensor. It is again
obtained using (\ref{met}),(\ref{metr}) and (\ref{duality}) and has
the form\tiny $$\label{dualmetriXtoalpha} \widetilde F_{ij}(\tilde
x)=
   \left(
\begin{array}{cccc}
 -\frac{b^2 \left({\tilde x_2}^2 (\alpha -2)^2 b^4+4\right)}{b^4 {\tilde x_1} (\alpha +2) (2 {\tilde x_2}
   (\alpha -2)+{\tilde x_1} (\alpha +2))-4} & \frac{4 b^2-b^6 {\tilde x_1} {\tilde x_2} \left(\alpha
   ^2-4\right)}{b^4 {\tilde x_1} (\alpha +2) (2 {\tilde x_2} (\alpha -2)+{\tilde x_1} (\alpha +2))-4} & 0
   & \frac{b^4 ({\tilde x_2} (\alpha -2)+{\tilde x_1} (\alpha +2))}{b^4 {\tilde x_1} (\alpha +2) (2
   {\tilde x_2} (\alpha -2)+{\tilde x_1} (\alpha +2))-4} \\
 \frac{4 b^2-b^6 {\tilde x_1} {\tilde x_2} \left(\alpha ^2-4\right)}{b^4 {\tilde x_1} (\alpha +2) (2
   {\tilde x_2} (\alpha -2)+{\tilde x_1} (\alpha +2))-4} & -\frac{b^6 {\tilde x_1}^2 (\alpha +2)^2}{b^4
   {\tilde x_1} (\alpha +2) (2 {\tilde x
   _2} (\alpha -2)+{\tilde x_1} (\alpha +2))-4} & 0 & -\frac{b^4
   {\tilde x
   _1} (\alpha +2)}{b^4 {\tilde x_1} (\alpha +2) (2 {\tilde x_2} (\alpha -2)+{\tilde x_1} (\alpha
   +2))-4} \\
 0 & 0 & -b^2 & 0 \\
 -\frac{b^4 ({\tilde x_2} (\alpha -2)+{\tilde x_1} (\alpha +2))}{b^4 {\tilde x_1} (\alpha +2) (2 {\tilde x_2}
   (\alpha -2)+{\tilde x_1} (\alpha +2))-4} & \frac{b^4 {\tilde x_1} (\alpha +2)}{b^4 {\tilde x_1} (\alpha
   +2) (2 {\tilde x_2} (\alpha -2)+{\tilde x_1} (\alpha +2))-4} & 0 & \frac{b^2}{b^4 {\tilde x_1} (\alpha
   +2) (2 {\tilde x_2} (\alpha -2)+{\tilde x_1} (\alpha +2))-4}
\end{array}
\right).
$$
\normalsize This tensor has nonzero and  nonconstant Gauss curvature
and torsion.
\subsection {$B$--field}
The \dd {} construction enables to add the $B$--field (torsion
potential) to the metric so that the resulting tensor $G'=G+B$ is
invariant with respect to the same isometry group as the metric
itself. Namely, changing $E_0$ to\footnote{Other antisymmetric
elements do not change torsion}
\begin{equation}\label{E0Xtoalphaasym}
    E_0=\left(
\begin{array}{cccc}
 0 & -\frac{1}{b^2}+\beta_1 & \beta_2& 0 \\
 -\frac{1}{b^2}-\beta_1 & -\frac{1}{b^2} & \beta_3 & 0 \\
 -\beta_2 & -\beta_3 & -\frac{1}{b^2} & 0 \\
 0 & 0 & 0 & -\frac{4}{b^2}
\end{array}
\right)
\end{equation}
and applying the formula (\ref{met}),  (\ref{metr}),  we get
covariant tensor that after the transformation
(\ref{tfnCoorsXtoalpha}) acquires the form
\begin{equation}\label{fantisym}G'_{ij}(u,v,x,y)=\left(
\begin{array}{cccc}
 -\frac{x^{\alpha -2}}{b^2} & -\frac{{\beta_1} b^2+1}{b^2 x^2} & 0 & x^{\frac{\alpha }{2}-2} {\beta_3} \\
 \frac{b^2 {\beta_1}-1}{b^2 x^2} & 0 & 0 & x^{-\frac{\alpha }{2}-2} {\beta_2} \\
 0 & 0 & -\frac{1}{b^2 x^2} & 0 \\
 -x^{\frac{\alpha }{2}-2} {\beta_3} & -x^{-\frac{\alpha }{2}-2} {\beta_2} & 0 & -\frac{1}{b^2 x^2}
\end{array}
\right).\end{equation} Its symmetric part is the metric
(\ref{metricXtoalpha}). This tensor is again invariant with respect
to the isometry group generated by
$K_I,\,K_{III},\,K_{IV},\,K_{VI}$. For $\beta_1=\beta_2=0$ the
invariant group can be extended by the generator $K_V$.

Torsion $H=dB$ obtained from the antisymmetric part of $G'$ is
\begin{equation}\label{torsionXtoalpha}
H=-4\beta_1\,du\wedge dv\wedge dy -\beta_2(4+\alpha)\,du\wedge
dx\wedge dy
 -\beta_3(4-\alpha)\,dv\wedge dx\wedge dy
\end{equation}
As the tensor (\ref{fantisym}) was obtained by the \dd {}
construction it is possible to dualize it but the result is too
extensive to display.

\section{Conclusion}Isometry groups of metrics can be used for construction of their
(nonabelian) T-dual backgrounds. Sufficient condition for that is
that the metric have an isometry subgroup whose dimension is equal
to the dimension of the Riemannian manifold and its action on the
manifold is transitive and free.

We have shown that for the plane wave metrics (\ref{FF}) and
(\ref{metricXtoalpha}) such isometry subgroups exist and the metrics
can be dualized by the \pl{} T--duality transformation. We have
determined the metrics and B--fields dual to the plane waves. For
homogeneous plane waves (\ref{FF}) we have also found the dilaton
field that guarantees conformal invariance of the dual metric.

Metrics that possess isometry group whose dimension is greater than
the dimension of the Riemannian manifold may have several duals.
More precisely, if the metric admits various isometry subgroups with
above given properties then we can construct several backgrounds
dual to the metric. This phenomenon is another kind of plurality of
sigma models different from the \pl{} T--plurality described in the
Section \ref{secPLT}.

An example of this type of plurality is provided by the plane wave
metric (\ref{FF}) with isometry subgroups generated by Killing
vectors $(K_{1},K_{2},K_{3},K_{6}+\rho K_7)$ or by
$(K_{1},K_{2},K_{5},K_{6})$ (see (\ref{killingshomogeneous})
producing two dual backgrounds (\ref{dualhomogenoeous}) and
(\ref{dualhomogenoeous2}). To decide if this plurality is different
from the \pl{} T--plurality one has to check whether the
eight--dimensional \dd s generated by the four--dimensional abelian
algebra and algebras spanned by $(K_{1},K_{2},K_{3},K_{6}+\rho K_7)$
or $(K_{1},K_{2},K_{5},K_{6})$ are isomorphic by a transformation
that leave the constant matrix (\ref{E0homogeneous}) invariant. This
is, however, very difficult task that might be investigated in the
future.
\section*{Acknowledgement}
This work was supported by the research plan LC527 of the Ministry
of Education of the Czech Republic. Consultation with P. Winternitz
and L. \v Snobl on classification of subalgebras are gratefully
acknowledged.


\begin{thebibliography}{999}
\bibitem{klise}{C. Klim\v c\'ik, P. \v Severa, \emph{Dual non-Abelian duality and the Drinfeld double},
Phys. Lett. B 351 (1995) 445.}
\bibitem{vall:su2}
M.A. Lledo and V.S. Varadarajan, \emph{{\rm SU}(2) Poisson-Lie
T-duality}, Lett. Math. Phys. {45} (1998) {247}, [hep-th {9803175}].

\bibitem{sfe:pltd}
K.~Sfetsos, \emph{Poisson-Lie T-duality beyond the classical level
and
  the renormalization group}, Phys.Lett B {432} (1998) {365}, [hep-th {9803019}].

\bibitem{hlas}{L. Hlavat\'y, L. \v Snobl, \emph{Poisson-Lie T-plurality of three-dimensional conformally invariant sigma models II: Nondiagonal metrics and dilatonpuzzle}, JHEP 0410:045, (2004).}
\bibitem{papa}{G. Papadopoulos, J.G. Russo and A.A. Tseytlin, \emph{Solvable model of strings in a time-dependent plane-wave background}, Class. Quant. Grav. 20:969-1016,(2003), [hep-th/0211289].}
\bibitem{blau}{M. Blau, M. O'Loughlin, \emph{Homogeneous Plane Waves}, Nucl.Phys. B 654 (2003) 135-176, [hep-th/021135].}
\bibitem{siklos}{S.T.C.Siklos, \emph{Lobatchewski plane gravitational waves} in Galaxies, Axisymmetric Systems and Relativity ed M A H MacCallum
(Cambridge: Cambridge University Press) 1985, p. 247.}
\bibitem{podolsky} {J.Podolsk\'y, \emph{Interpretation of the Siklos solutions as exact gravitational
waves in the anti-de Sitter universe}, Class. Quantum Grav. 15.
(1998) 719-733}.
\bibitem{kli:pltd}
C.~Klim\v{c}\'{\i}k, \emph{Poisson-Lie T-duality}, Nucl. Phys. {46}
(Proc.Suppl.) (1996) {116},  [hepth{9509095}].
\bibitem{unge:pltp}
R.~von~Unge, \emph{Poisson-Lie T-plurality}, J. High Energy Phys.
{07} (2002) {014}, [hepth{0205245}].
\bibitem{kaigor}{Kaigorodov V R, \emph{Einstein spaces of maximum mobility},  Dokl. Akad. Nauk. SSSR 146 (1962) 793; Sov. Phys. Doklady 7 (1963)
893}.
\bibitem{winter}J. Patera, P. Winternitz, and H. Zassenhaus,
\emph{Continuous subgroups of the fundamental groups of physics. I.
General method and the Poincar\'e group}, J. Math. Phys. 16 (1975)
1597-1614.

\end{thebibliography}
\end{document}